\documentclass[twocolumn,preprintnumbers,aps,showpacs]{revtex4}

\def\bea{\begin{eqnarray}}
\def\eea{\end{eqnarray}}
\def\ben{\begin{equation}}
\def\een{\end{equation}}
\def\benu{\begin{enumerate}}
\def\enu{\end{enumerate}}

\def\sss{\scriptscriptstyle\rm}

\def\1var{(\bx_1...\bx\N)}

\def\br{{\bf r}}

\def\bx{{x}}

\def\x{_{\sss X}}

\def\s{_{\sss S}}
\def\xc{_{\sss XC}}
\def\Hxc{_{\sss HXC}}

\def\N{_{\sss N}}
\def\H{_{\sss H}}

\def\ee{_{\rm ee}}

\def\sph_int{ {\int d^3 r}}

\usepackage{graphicx}
\usepackage{psfrag}
\usepackage{amsmath}
\usepackage{bm}
 
\input epsf

\begin{document}
\title{Long-range excitations in time-dependent density functional theory}
\date{\today}
\author{Neepa T. Maitra and David G. Tempel}
\affiliation{Department of Physics and Astronomy, Hunter College and City University of New York, 695 Park Avenue, New York, NY 10021, USA}

\begin{abstract}
Adiabatic time-dependent density functional theory fails for
excitations of a heteroatomic molecule composed of two open-shell
fragments at large separation.  Strong frequency-dependence of the
exchange-correlation kernel is necessary for both local and
charge-transfer excitations. The root of this is static correlation
created by the step in the exact Kohn-Sham ground-state potential
between the two fragments.  An approximate non-empirical kernel is
derived for excited molecular dissociation curves at large
separation. Our result is also relevant for the usual local and
semi-local approximations for the ground-state potential, as static
correlation there arises from the coalescence of the highest occupied
and lowest unoccupied orbital energies as the molecule dissociates.
\end{abstract}
\pacs{31.15.Ew,31.10.+z,31.25.Jf}
\maketitle 

\section{Introduction and Motivation}
Time-dependent density functional theory (TDDFT) has seen a surge of
applications in recent years~\cite{RG84,PGG96,C96,MBAG02,MG04,WBG05}, with
calculations of electronic excitations in a wide variety of systems of
physical, chemical, and biochemical interest. 
Errors in TDDFT excitation energies are typically within a few tenths of an eV, while implementations can scale with size even more favorably than that of time-dependent Hartree-Fock. 
In principle, TDDFT yields {\it exact} excitation spectra: the errors
are due to approximations to the exchange-correlation
(xc) functionals.  Approximations enter at two stages in usual linear
response calculations: first, in the xc contribution to the one-body
ground-state Kohn-Sham (KS) potential out of which bare excitations
are calculated. For example, its too rapid asymptotic decay in LDA/GGA
causes problems for high-lying bound states~\cite{CJCS98,TH98b,WMB03,WB05}.
 Second, the dynamic xc kernel must be approximated; this corrects the bare KS transitions
towards the true transitions. Almost all calculations employ an
adiabatic approximation, which entirely neglects 
frequency-dependence in the kernel. Its  remarkable success for many
excitations is not well understood. Understanding where it fails is important in constructing more accurate
functional approximations. 
The lack of frequency-dependence is why adiabatic TDDFT (ATDDFT) fails for states of double
excitation character~\cite{MZCB04,CZMB04,C05,TH00,HH99}.

Molecular dynamics simulations, atom trapping, and photoassociation
spectroscopy all need accurate calculations of excited molecular
dissociation
curves~\cite{CGGG00,GGGB00,FR05,CCS98,WGBN04,LKQM05,fromundul}, but these are challenging for approximate methods. Even ground-state dissociation is difficult~\cite{FNGB05}. Closely linked are the
difficulties in obtaining long-range charge-transfer (CT)
excitations. The latter are notoriously underestimated in
ATDDFT~\cite{DWH03,DH04,T03,GB04b,TAHR99,CGGG00,GGGB00,TTYY04,WV05,M05c}, yet
are vital in biological and chemical systems large
enough that density functional methods are the only feasible
approach. 

We show here that a
strong frequency-dependence in the kernel is needed near {\it every}
excitation of a  molecule composed of two different open-shell
fragments at large separation: The ubiquitous
adiabatic approximation fails for {\it all} excitations.
We show this is due to  static correlation in the KS
ground-state, leading to every single  excitation being almost
degenerate with a double excitation. 
We resolve what the frequency-dependence of the exact xc kernel is for the tails of excited dissociation curves, and derive an approximation based
on this.
Our kernel yields both charge-transfer and local
excitations, and contains the correct dissociation limit and
leading-order polarization. It bootstraps an
adiabatic approximation, and yields results to the same good level of
accuracy that ATDDFT has for more common local excitations: that
is, our kernel undoes the problematic static correlation in the KS
system.


A crucial role is played by the step structure in the exact ground-state KS potential
that develops as a heteroatomic molecule dissociates
~\cite{PPLB82,P85b,AB85b,PL97c,AB85}. The step has size equal to the
difference in the ionization potentials of the two fragments, so
aligns the highest occupied molecular orbitals (HOMOs) of each.  Its
consequences for TDDFT have only begun to be
explored~\cite{M05c,LK05}.  Recently we showed that near the {\it
lowest} CT excitation on each fragment, the step imposes a strong
frequency-dependence on the exact kernel~\cite{M05c,erratum}.
Here, we go much further, showing the dramatic frequency-dependence
required for both higher CT {\it and} all local excitations.
 Local
and CT excitations are intimately entangled by the step in the KS
system and we construct the kernel that disentangles them. 

It is important to note that although local and semi-local
approximations to the ground-state potential (ie LDA and GGAs) lack
the step, static correlation still features, since in these
cases, the HOMO and lowest unoccupied molecular orbital (LUMO) become
degenerate as the molecule dissociates. So, the construction of our xc
kernel has relevant implications for the case of the usual approximate
ground-state KS potentials too.

In Sec.~\ref{sec:TDDFTLinearResponse}, we briefly review linear
response theory in TDDFT. We discuss its application to CT states of a
long-range molecule and point out a fundamental difference in the
analysis when the closed-shell molecule consists of open-shell
fragments rather than closed-shell fragments. We summarize the recent
result of Ref.~\cite{M05c} on the structure of the exact kernel that
captures the lowest CT excitations exactly in the case of open-shell
fragments in the large-separation limit.  In
Sec.~\ref{sec:LocalandHigher} we extend the analysis to all
excitations of the molecule, showing how both local and CT excitations
require strong frequency-dependence in the xc kernel that is lacking
in all approximations to date. We derive an approximation to
the kernel and demonstrate it on a simple molecule. 
Finally,
Sec.~\ref{sec:Outlook} contains a discussion on implications of our
result.



\section{TDDFT Linear Response and Lowest Charge-Transfer States}
\label{sec:TDDFTLinearResponse}
TDDFT linear response~\cite{PGG96,C96} proceeds by correcting
the KS single excitations towards the true excitations, (which may be
mixtures of any number of excitations) through the
Hartree-exchange-correlation kernel $f\Hxc[n_0](\br,\br',\omega) =
1/\vert \br - \br'\vert + f\xc[n_0](\br,\br',\omega)$,
a functional of
the ground-state density $n_0(\br)$. The xc kernel is the
frequency-time Fourier-transform of 
the functional derivative of the
xc potential, 
$f\xc[n_0](\br,\br',t-t') = \delta v\xc(\br t)/\delta n(\br' t)\vert_{n_0}$. 
The density-density response function of the KS system $\chi\s$ and that of the interacting system $\chi$ are related through:
\ben
\hat\chi^{-1}(\omega)= \hat\chi\s^{-1}(\omega) - \hat f\Hxc(\omega)\,.
\label{eq:dyson}
\een 
Usually a matrix version~\cite{C96} is used to obtain excitation
energies and oscillator strengths. The vast majority of calculations
utilize an adiabatic approximation for the xc kernel, i.e. one that is
frequency-independent.  Typically the results are accurate to within a
few tenths of an eV (although there are exceptions, see eg. ~\cite{TH98b,HH99,CZMB04,DWH03}).

When excitations are well-separated, the matrix may be simplified in a
``single pole approximation (SPA)''~\cite{VOC99,GPG00,AGB03} that
gives the true frequency $\omega$ as a shift from a KS transition: for
a closed-shell molecule,
\ben 
\omega = \omega_q +2[q|f_{\sss HXC}(\omega_q)|q]\,.
\label{eq:spa}
\een
Here, $q = (i,a)$ is a double-index labeling the single KS excitation from an occupied orbital $\phi_i$ to unoccupied $\phi_a$, $\omega_q= \epsilon_a - \epsilon_i$, and 
$[q|f_{\sss HXC}(\omega)|q] = \int d\br d\br' \phi_i^*(\br)\phi_a(\br) f\Hxc(\br,\br',\omega) \phi_{i}(\br')\phi_{a}^*(\br')$.

The SPA is also an important tool in understanding the TDDFT
corrections to the bare KS energies. Applying Eq.~(\ref{eq:spa}) to the problem of
lowest charge-transfer states in a long-range molecule~\cite{DWH03},
shows immediately that the correction from the kernel, $[q|f_{\sss HXC}(\omega_q)|q]$, vanishes,
because there is vanishing overlap between the occupied orbital
$\phi_i(\br)$ on the donor and the unoccupied $\phi_a(\br)$ on the
acceptor, in the limit of their large separation $R$. The TDDFT value of Eq.~(\ref{eq:spa}) then reduces to the bare KS
eigenvalue difference, i.e. the acceptor LUMO orbital energy  minus the donor HOMO orbital energy. The latter is equal to minus the ionization energy of
the donor, $I_D$, while the former is the KS electron
affinity\cite{P85b, PL97c,PPLB82,AB85}, $A_{A, \sss S} = A_A - A_{A, \sss XC}$, not the true electron
affinity of the acceptor, $A_A$. The resulting frequency, $\omega = I_D - A_{A,\sss S}$, is a severe underestimate to the true CT energy:
aside from the usual local/semi-local approximations underestimating the
ionization energy, it lacks the xc contribution to the electron
affinity, as well as the $-1/R$ tail~\cite{DWH03,DH04,T03,GB04b}.  The exact CT energy is $I_D - A_A -1/R$ in the large-R limit.


When the (closed-shell) long-range molecule consists of two open-shell
fragments~\cite{M05c} (labelled as 1 and 2 below), this argument needs to be revisited because
neither the HOMO nor LUMO are localized on one of the two
fragments. In the large separation limit, the exact KS potential
contains a step~\cite{P85b,AB85,GB96} that exactly aligns the
individual HOMO's of the fragments,  rendering the HOMO
of the long-range molecule, $\phi_0$, to be the bonding molecular orbital
composed of the HOMO's of the individual fragments, $\phi_0 = (\phi_{H,1} +\phi_{H,2})/\sqrt{2}$. The LUMO $\overline\phi_0$ is the anti-bonding
molecular orbital. (See Fig.1B, and also e.g. Fig. 1 of
Ref.~\cite{M05c}, Fig. 7 of Ref.~\cite{P85b}, and Fig. 1 of
Ref.~\cite{GB96}).  
The lowest CT states lie in the subspace formed by the HOMO and LUMO, as shown by the explicit diagonalization of the Hamiltonian in Ref.~\cite{M05c}.
Turning to TDDFT, Eq.~(\ref{eq:spa}), the
bare KS energy difference, $\omega_q$, vanishes as an exponential
function $e^{-R}$ of the separation of the two fragments, $R$, since
the antibonding LUMO and bonding HOMO are separated in energy only by
the tunnel splitting through the barrier created by the step.  The
correction from the xc kernel in Eq.~\ref{eq:spa} is no longer zero,
as these orbitals have significant overlap in both the atomic
regions. However any {\it adiabatic} approximation to the kernel will 
lead to drastically incorrect CT energies: The underlying reason is because the
adiabatic approximation entirely neglects
double-excitations~\cite{MZCB04,CZMB04}, which are also absent in the non-interacting response function $\chi\s$. Yet the double excitation to the
antibonding orbital is critical in this case to capture the correct
nature of the true excitations in this subspace~\cite{M05c} (as well
as the Heitler-London ground-state). Refs.~\cite{MZCB04,CZMB04} showed how the mixing of a double excitation with a single can be incorporated in TDDFT via a {\it dressed SPA}, $\omega = \omega_q +2[q\vert f\Hxc(\omega)\vert q]$: here the kernel has strong frequency-dependence. 
In Ref.~\cite{M05c} the form of the exact non-adiabatic xc kernel was 
uncovered in
the limit of large separation and in the limit that the CT excitations have
negligible coupling to all other excitations of the system. This was
obtained by examining the structure of the response functions $\chi$
and $\chi\s$ in this subspace and comparing with the explicit
diagonalization of the Hamiltonian. The result was:
\ben
\bar\omega[q\vert f\Hxc(\omega)\vert q] = \delta^2 + \frac{\omega_1\omega_2 -\bar{\omega}^2}{4} + \frac{\omega_1\omega_2\delta^2}{\omega^2 - \omega_1\omega_2}
\label{eq:result1st}
\een 
for the KS transition $q=$bonding HOMO$\to$anti-bonding LUMO orbital.
Here, 
$\omega_1=I_2 - A_1 - 1/R$ and $\omega_2=I_1 - A_2 - 1/R$,
where $A = A +A\xc^{\sss approx}$, with the xc contribution to the electron
affinity of each fragment approximated as $A\xc^{\sss approx} = -\int d^3r\int
d^3r' \phi_H(\br)^2\phi_H(\br')^2/\vert\br - \br'\vert$, and $\phi_H$ is
the HOMO of the appropriate fragment. Finally,
$\delta= (\omega_1 -\omega_2)/2$, and $\bar\omega \sim e^{-R}$ is the HOMO-LUMO splitting (the bare KS frequency)~\cite{M05c, erratum}.
 Notice the strong non-adiabaticity in the exact kernel, manifest by
the pole in the denominator of Eq.~(\ref{eq:result1st}).

The strong-frequency dependence in this expression arises due to
static correlation in the system: the near-degeneracy of the
antibonding orbital leads to the KS ground-state Slater determinant
being near-degenerate to two others. This feature makes heteroatomic
dissociation in the ground and lowest CT states resemble homoatomic
dissociation in some respects~\cite{GGGB00}; problems associated with
ground-state homoatomic dissociation are now well-known~\cite{PSB95}.
In the present paper we explore the consequences of the static
correlation on {\it all} other excitations of the {\it heteroatomic}
system, and model the required frequency-dependence in the kernel.
Every single excitation out of the occupied bonding orbital is almost
degenerate with a double excitation, where another electron is excited
from the bonding orbital to the antibonding orbital . This
near-degeneracy leads to strong-frequency dependence in the xc kernel
relevant to all excitations of the molecule.

It is very interesting to note that although the usual
local/semi-local approximations (LDA/GGA's) to the ground-state
potential do not contain the step, the HOMO and LUMO are nevertheless
also delocalized over the long-range molecule. We shall return in
Sec.~\ref{sec:Outlook} to a discussion of this.

\section{Local and Higher Charge Transfer Excitations}
\label{sec:LocalandHigher}
\begin{figure}[h]
 \centering 
 \includegraphics[height=5.5cm,width=7cm]{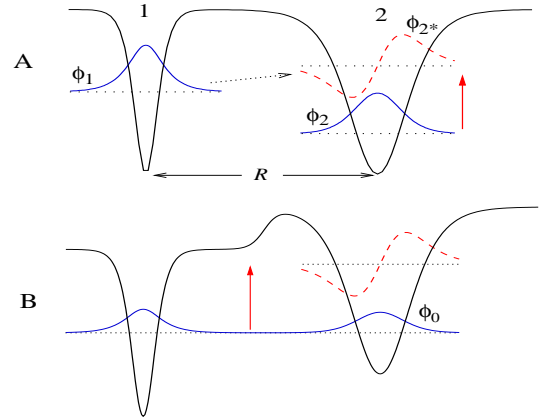} 
\caption{A: Model of long-range molecule. The Heitler-London singlet ground-state is $\left(\phi_1(\br)\phi_2(\br') + \phi_2(\br)\phi_1(\br')\right)/\sqrt{2}$. The dashed curve is an excitation on ``atom'' 2. B: KS potential, (which eventually returns to zero on the right). The KS ground-state is the doubly-occupied bonding orbital $\phi_0$.}
\label{fig:exc2}
\end{figure}

Consider first a model molecule composed of two different ``one-electron
atoms''. At large separations, the true Heitler-London ground-state (Fig.~\ref{fig:exc2}A), has one electron on
each atom (in orbitals $\phi_1$ and $\phi_2$ respectively).
On the other hand, due to the alignment of the atomic
levels caused by the step in the KS potential
(Fig.~\ref{fig:exc2}B), the KS ground-state is the doubly-occupied
bonding orbital, $\phi_0 = (\phi_1 + \phi_2)/\sqrt{2}$, with each electron  evenly spread over each atom.
Generically, higher excitations of the atoms do not coincide;
for our example, atom 2
has one higher bound excitation, whereas atom 1 has none.

The vertical arrow in Fig.~\ref{fig:exc2}A illustrates a local excitation on
atom 2, $\phi_2 \to \phi_{2^*}$.  If we attempt to describe this in
the KS system, we immediately see a problem: Excitations must occur out of
the occupied bonding orbital $\phi_0$ (Fig.~\ref{fig:exc2}B), so
excitation leaves the molecule with ``half an electron''
on atom 1 and ``one and a half electrons'' on atom 2. This single
excitation, $\phi_0 \to \phi_{2^*}$ is a poor description of the
excitation of the true system. The true excitation is in fact a
linear combination of this single excitation and a double
excitation, $(\phi_0,\phi_0) \to (\phi_{2^*},\overline{\phi_0})$ where
the other electron occupying the bonding orbital is excited to the
antibonding orbital $\overline{\phi_0}$ .  (We assume for now the other
excitations of the system are farther away in energy). 
Because the
transition frequency to the antibonding orbital is the tunnel
frequency between the atoms, it is exponentially small as a function of
their separation. Thus, {\it every} single excitation of the
system $(\phi_0 \to \phi_a)$ is almost degenerate with the double
excitation $(\phi_0,\phi_0) \to (\phi_{a},\overline{\phi_0})$. This feature is a signature of static correlation in the KS ground state. Since
double excitations are missing in the KS linear
response~\cite{MZCB04}, it is the job of the xc kernel to fold them
in (see shortly).

It is instructive first to diagonalize the interacting Hamiltonian $H$ in the $2 \times 2$ KS basis, whose KS energies differ only by the tunnel splitting($\sim e^{-R}$):
\bea
\nonumber
\Phi_q &=& (\phi_0(\br) \phi_{2^*}(\br') +\phi_{2^*}(\br) \phi_0(\br'))/\sqrt{2}  \\
\Phi_D &=& (\overline{\phi_0}(\br) \phi_{2^*}(\br') +\phi_{2^*}(\br) \overline{\phi_0}(\br'))/\sqrt{2}
\label{eq:KSbasis}
\eea
$H$ sums the kinetic,
external potential, and electron interaction $(V\ee)$ operators, and rotates the pair into:
\bea
\nonumber
\Psi_{2^*} &=& \left(\phi_1(\br)\phi_{2^*}(\br') + \phi_{2^*}(\br)\phi_1(\br')\right)/\sqrt{2}\\
\Psi_{CT 2^*} &=&\left(\phi_2(\br)\phi_{2^*}(\br') + \phi_{2^*}(\br)\phi_2(\br')\right)/\sqrt{2}  
\label{eq:2x2}
\eea
Within the truncated space, Eqs.~(\ref{eq:2x2}) are eigenstates of the
interacting system: $\Psi_{2^*}$ is a local excitation on atom 2,
while $\Psi_{CT 2^*}$ is an excited CT from atom 1 to an excited state
of atom 2 (dotted arrow in Fig.~\ref{fig:exc2}A).  Because these
states are paired together due to the bonding orbital nature of the KS
ground state, the two qualititatively different excitations appear
together in the exact TDDFT, as we will see shortly.  The eigenvalues of the diagonalization give their
approximated frequencies,
after subtracting the expectation value of $H$ in the Heitler-London
ground-state, $E_{HL}$, (see also Refs.~\cite{M05c,MZCB04}):
\bea
\nonumber
\omega_{2^*}&=& \epsilon_{2^*} - \epsilon_2 \\
\omega_{CT 2^*}&=& \epsilon_{2^*} - \epsilon_1 - A\x^{(2^*)} -1/R\,,
\label{eq:naive}
\eea
to leading order in $R$, 
where  
\bea
\nonumber
A\x^{(2^*)}& =& -\int d\br \int d\br' \left(\vert \phi_{2^*}(\br)\vert^2 \vert \phi_{2}(\br')\vert^2 \right.\\
&+&\left.\phi_{2^*}^*(\br)\phi_2^*(\br')\phi_{2^*}(\br')\phi_2(\br)\right)V\ee(\br - \br')
\label{eq:Ax*}
\eea
This approximates the xc part of the excited electron affinity $A^{(2*)}$: generally for an $N$-electron species,
\ben
A^*  = E_0(N) - E^*(N+1)= A\s^* + A\xc^* 
\label{eq:Axc*}
\een
with $A\s^* =-\epsilon^*$, the energy of the excited KS orbital that the added electron joins.  $E_0(N)$ is the ground-state energy of the $N$-electron system, and $E^*(N+1)$ is the energy of the excited state of the $(N+1)$-electron system.
The excited xc electron affinity accounts for relaxation 
when an electron is added to an $N$-electron system, forming an $(N+1)$-electron excited state. 
  (No correction is needed for ionization out of species 1~\cite{PL97c}: the KS HOMO energy exactly equals minus the ionization potential.)

Eqs.~(\ref{eq:naive}) and~(\ref{eq:Ax*}) become exact in the weak
interaction limit where there is no coupling to  other KS excitations. Dynamic correlation is completely missing.
Although derived for two one-electron atoms, the case for $N_{1,(2)}$ electrons on fragment 1(2) (both neutral and open-shell) follows analogously.
Eq.~(\ref{eq:Ax*}) is an exchange approximation which may also be obtained 
from perturbation theory: One computes the $V\ee$ matrix element
in an $(N_2+1)$ electron Slater determinant, where the added electron is
placed in an unrelaxed virtual orbital of the $N_2$-electron KS potential.

\begin{figure}
 \centering 
 \includegraphics[height=4.5cm,width=7cm]{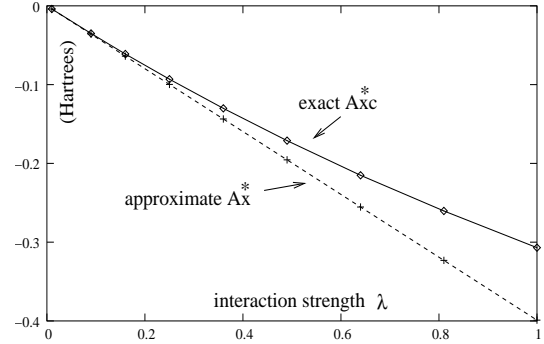} 
\caption{Exchange-correlation contribution to the excited electron affinity for a test system (See text). }
\label{fig:Axcplot}
\end{figure}

Before turning to TDDFT, we first give a simple example to test Eq.~(\ref{eq:Ax*}). We consider one fermion in a one-dimensional parabolic well~\cite{MZCB04}, $H = -\frac{1}{2}\frac{d^2}{dx^2} + \frac{1}{2}x^2$. 
Excited
states of two fermions in the well, interacting via a scaled delta-interaction, $V\ee=\lambda\delta(x - x')$, may be 
 numerically calculated at all interaction strengths $\lambda$~\cite{MZCB04}.  Figure~\ref{fig:Axcplot} compares
the affinity calculated from energy differences (Eq.~\ref{eq:Axc*}), with Eq.~(\ref{eq:Ax*}): as expected, Eq.~(\ref{eq:Ax*}) is
exact for weak interaction.

Returning to TDDFT, the diagonalization process above is effectively
hidden in the structure of the xc kernel. Our approximation for $f\xc$
is motivated by the above analysis and also by the form of the
interacting and KS density-density response functions.

Because only single
excitations appear in the KS response, in a SPA $\chi\s$ has one pole at $\omega_q = \epsilon_{2^*} - \epsilon_2$ ($\Phi_q$ of Eq.~\ref{eq:KSbasis}). The true density-density response however has two poles, one at the local excitation $\omega_{2^*}$, and the other at the CT excitation $\omega_{CT 2^*}$. Generating an extra pole in Eq.~(\ref{eq:dyson}) and folding in the double excitation $(\phi_0,\phi_0) \to (\phi_{2^*},\overline{\phi_0})$, requires a dressed (i.e. frequency-dependent) SPA (see also Ref.~\cite{MZCB04}), with the form
$[q|f_{\sss HXC}(\omega)|q] = a + b/(\omega -c)$.
Note here that $\vert q] = \phi_0(\br)\phi_{2^*}(\br)$. 
This form is consistent with the earlier diagonalization analysis (yielding Eqs.~\ref{eq:naive}),
 which may be written
\ben
\omega = \langle\Phi_q\vert H\vert\Phi_q\rangle - E_{HL} + \frac{\vert\langle\Phi_q\vert H\vert\Phi_D\rangle\vert^2}{\omega  -(\langle\Phi_D\vert H\vert\Phi_D\rangle - E_{HL})}\,,
\label{eq:diag}
\een
 A {\it first} approximation
for the $f\xc$ matrix element then results directly from 
subtracting $\omega_q = \epsilon_{2^*} - \epsilon_2$ from the right-hand-side of Eq.~(\ref{eq:diag}) (c.f. Eq.~(\ref{eq:spa})).
Although this would give a huge improvement over any adiabatic
approximation (see later),
it
lacks the correct local Hartree-exchange-correlation effects. 
We now modify the approximation to incorporate these. 

As $\langle\Phi_D\vert H\vert\Phi_D\rangle = \langle\Phi_q\vert H\vert\Phi_q\rangle + O(e^{-R})$ at large $R$, Eq.~(\ref{eq:diag}) suggests a two-parameter
model,
\ben
2[q\vert f\Hxc(\omega)\vert q] = a - \omega_q + \frac{b^2}{\omega - a} \,. 
\label{eq:model}
\een
We set these parameters by requiring the solutions of Eq.~(\ref{eq:spa}) with Eq.~\ref{eq:model}, i.e. $\omega = a
\pm b$, reproduce ATDDFT values for local xc effects and for lowest order polarization. That is, 
\ben
a=\frac{1}{2}(\omega_{2^*} + \omega_{CT 2^*})\,,\,\, b = \frac{1}{2}(\omega_{2^*} - \omega_{CT 2^*}),
\een
where
\bea
\nonumber
\omega_{2^*}&=&\omega_q +\Delta_{2^*} + D(1,2^*) - D(1,2)\,, {\rm and}\\
\nonumber
\omega_{CT 2^*}&=& \omega_q +\Delta_{CT 2^*} - \frac{1}{R} + D(1^+,(2^-)^*) - D(1,2) \\
\label{eq:parameters}
\eea
(Note again $\omega_q = \epsilon_{2^*} -\epsilon_2$ is the bare KS frequency). 
Here ATDDFT calculations on separated fragments determine all the quantities:
(i) $\Delta_{2^*}$ is the ATDDFT value for the excitation on  fragment 2, minus the KS frequency $\epsilon_{2^*} -\epsilon_2$. 
For example, in a SPA, $\Delta_{2^*} = 2[q\vert f\H +f_{{\sss XC},\uparrow\uparrow}^A\vert q]$, where $f_{{\sss XC},\uparrow\uparrow}^A = \delta^2 E\xc/\delta n_\uparrow(\br)\delta n_\uparrow(\br')$, with $E\xc$ being a ground-state xc energy approximation. 
\newline
(ii)  $\Delta_{CT 2^*} = -(I_2 - I_1) - A\xc^{(2*)}$, where we write $A\xc^{(2^*)} = A(N_2) - \omega^*(N_2+1) +\epsilon_{2^*}$. Here $A(N_2)$ is the usual electron affinity of fragment 2, that may be obtained from ground-state energy differences between  the negative ion formed by adding an electron to fragment 2, and that of the neutral fragment 2.  The frequency of the excited state of the $(N_2 +1)$-electron ion 2, $\omega^*(N_2+1)$, is then given by ATDDFT performed on this negative ion. 
\newline
(iii) $D(1,2)$ is the dipole-dipole energy between fragment 1 and fragment 2 in their ground states; $D(1,2^*)$ is that when fragment 2 is in the excited state; $D(1^+,(2^-)^*)$ is that between the positive donor 1 in its ground state and the excited acceptor state on fragment 2. The dipole moments can either be directly obtained from the ground-state DFT densities, or extracted from ATDDFT response on separated neutral or charged fragments.

Thus we obtain the dressed kernel matrix element Eq.~(\ref{eq:model}) in terms of KS quantities and ATDDFT run on the separated neutral and charged fragments.

As an example, consider two high lying excitations of the BeCl+
molecule, that dissociates in its ground state to Be$^+$ + Cl. We
consider the local excitation that dissociates to Be$^+$(3s) + Cl(3p)
and the CT excitation that is Be(3s) + Cl$^+$(3p).  We use the B3LYP
functional for the ATDDFT pieces of the calculation, as programmed in
the NWChem code~\cite{nwchem}.  Our approximate kernel yields the tail
of the Be(3s) + Cl$^+$(3p) CT dissociation curve to have the frequency
$(0.362 -1/R)$H, and the local Be$^+$(3s) excitation to be
$0.394$H. These have the correct dependence up to $O(1/R^3)$ on the
separation, and the asymptotes correspond to the generally reliable
ATDDFT values on the isolated Be+ and Cl fragments. On the other hand,
a purely adiabatic approximation applied to the molecule yields only
the local 2s$\to$ 3s excitation on Be+, with frequency $0.430$H; giving only half the usual
ATDDFT correction for a local excitation on an isolated fragment (see
also Sec.~\ref{sec:Outlook}), and completely missing the CT
one. Comparing with the atomic spectroscopic data of NIST for the
transition frequencies~\cite{nist}, we find the local excitation to be
$0.402$H and the CT asymptote to be $0.386$H.

\section{Discussion and Outlook}
\label{sec:Outlook}
Eq.~(\ref{eq:model}), together with the parameters of
Eq.~(\ref{eq:parameters}), is our approximation for the kernel. The
strong frequency-dependence is a consequence of static correlation in
the KS system: our result does not go beyond the accuracy that ATDDFT
has for usual local excitations, so it demonstrates the complications
static correlation creates for TDDFT. The static correlation is caused
by the near-degeneracy of the HOMO and LUMO orbitals due to the step
in the KS potential.  Our result provides a practical scheme to deal
with this, constructed from ATDDFT calculations run on the individual
fragments, and builds in first-order molecular polarization effects.
We note that our result is only accurate for the tail end of the
excited molecular dissocation curves (up to $O(1/R^3)$). No xc effects
across the long-range molecule are included.  Molecular Feshbach
resonances at frequencies higher than $I_1$ cannot be accurately
described; these appear as KS shape resonances tunneling through the
step.  Limited in this way, and yet having complicated form, our
approximation highlights challenges for a more complete description of
dissociation.

There is no exponential growth of the kernel
matrix element with separation, that one would expect from an 
{\it adiabatic} single pole analysis of long-range CT with {\it
localized} (atomic)orbitals~\cite{GB04b}; there the kernel would need
to grow as $e^R$ in order to get a non-vanishing xc affinity.
However, as shown here, for a long-range molecule consisting of
open-shell fragments, this does not occur with the correct {\it
delocalized} molecular orbitals, provided our {\it non-adiabatic}
approximation is used, as the CT state is ``carried along'' with the
local excitation, through KS excitation out of the bonding orbital.

We note that the lowest CT states in Ref.~\cite{M05c}, and discussed in Sec.~\ref{sec:TDDFTLinearResponse} here, are a distinct
case from the higher excitations considered here because there the
underlying KS transition was nearly-degenerate with the KS
ground-state itself: a single excitation to the antibonding transition
$\overline\phi_0$.  By perturbative arguments, the KS response goes as
the inverse frequency $\sim e^R$. The structure of the resulting
kernel (Eq.~(\ref{eq:result1st}) in the present paper)
is distinct from that for the higher local and CT excitations here
(Eqs.~\ref{eq:model} and \ref{eq:parameters}), but also displays
strong frequency-dependence.


Any adiabatic approximation is
stuck with unphysical half-electrons on each atom. The correction
to the bare KS energy would be half of what it should be for the local
excitation (since the $f\xc$ matrix element involves the bonding
rather than the localized orbital), and
entirely misses the CT one.

The step appears in the exact ground-state KS potential, and in
orbital-dependent approximations~\cite{G05}. If instead LDA, or a GGA,
was used for the ground-state potential, as in most calculations
today, there is no step. The HOMO and LUMO are nevertheless also
delocalized over the long-range molecule. It is now
well-established~\cite{PPLB82,P85b,AB85b} that the dissociation limit
of the molecule places (unphysical) fractional charges on each
open-shell fragment in approximations such as LDA. The individual
atomic KS potentials become distorted from their isolated form in such
a way as to align the individual HOMO's with each other, yielding
molecular orbitals that span the molecule, albeit unevenly.  The HOMO
and LUMO energies increasingly approach each other as the molecule
dissociates, yielding static correlation in the long-range molecule. Clearly, in light of the results here, an adiabatic
kernel used in conjunction with such a ground-state potential, will
fail for all excitations, and give unphysical fractional charges on
the dissociating fragments in the excited states.
Frequency-dependence is needed to fold in the double-excitation
associated with excitation into the LUMO.  Although the result for the
xc kernel of Ref.~\cite{M05c}, and our results in
Sec.~\ref{sec:LocalandHigher} are derived with the exact ground-state
potential in mind, it may be interesting to explore the use of our
kernel on top of an LDA ground-state potential, as much of the
essential physics is captured. The results of this paper will be an
important guide for likely modifications required in this case.



We thank the ACS Petroleum Research Fund and the National Science Foundation Career Program (CHE-0547913) for financial support, and Kieron Burke and Fan Zhang for discussions.


\begin{thebibliography}{99}
\bibitem{erratum}
Note a typo in Ref.~\cite{M05c}: the $\omega_{av}$ parameter should be replaced by $\delta = (\omega_1 -\omega_2)/2$, as appears in Eq.~(\ref{eq:result1st}) here.
\end{thebibliography}
\end{document}